\documentstyle[aps]{revtex}
\begin{document}
\draft
\title{Origin of entropy convergence in hydrophobic hydration and 
protein folding}
\author{
Shekhar Garde,$^{1,2}$
Gerhard Hummer,$^{1}$
Angel E. Garc{\'{\i}}a,$^1$
Michael E. Paulaitis,$^{2,3}$
and Lawrence R. Pratt$^1$}
\address{$^1$Theoretical Division, Los
Alamos National Laboratory, Los Alamos, NM 87545.}
\address{$^2$Center for Molecular and Engineering Thermodynamics,
Department of Chemical Engineering, \newline University of Delaware,
Newark, DE 19716} 
\address{$^3$ Department of Chemical
Engineering, Johns Hopkins University, Baltimore, MD 21218.}

\date{\today, LA-UR-96-2773}
\maketitle

\begin{abstract} 
An information theory model is used to construct a molecular
explanation why hydrophobic solvation entropies measured in
calorimetry of protein unfolding converge at a common temperature.
The entropy convergence follows from the weak temperature dependence
of occupancy fluctuations for molecular-scale volumes in water. The
macroscopic expression of the contrasting entropic behavior between
water and common organic solvents is the {\it relative\/} temperature
insensitivity of the water isothermal compressibility. The information
theory model provides a quantitative description of small molecule
hydration and predicts a {\it negative\/} entropy at convergence.
Interpretations of entropic contributions to protein folding should
account for this result.
\end{abstract}


High sensitivity calorimetry on the unfolding of globular proteins has
suggested that hydrophobic contributions to the entropies of unfolding
converge to zero near 385~K \cite
{Privalov:74,Privalov:79,Privalov:88,Privalov:93,Privalov:95}.
Additional thermodynamic information on protein folding processes is
then obtained by extrapolating the calorimetric measurements to the
convergence temperature where hydrophobic contributions vanish.  The
convergence behavior of entropies of solution of hydrocarbons in water
is known also \cite {Baldwin:86,Baldwin:92,Freire:92,Lee:91}.  Yet a
microscopic-level mechanism for this phenomenon has not been
offered\cite {Kauzmann:87}.  Here we identify a mechanism by analysis of
an information theory model of hydrophobic hydration \cite
{Hummer:96,berne:96}.  We show how the theory predicts entropy
convergence on the basis of the density and density fluctuations of
liquid water.  Consistent with experimental results on hydrophobic
hydration, the model predicts that the entropy at convergence should be
negative.

Aqueous protein solutions are complex systems involving molecular
interactions of several kinds, including those associated with ionic and
polar groups, in addition to hydrophobic contributions\cite
{Privalov:95}.  For clarity we focus on strictly hydrophobic species and
idealize those solutes as hard core objects that perfectly repel water
molecule centers identified as the position of the oxygen atoms.
Statistical mechanics relates the excess chemical potential of hard core
solutes to the probability, $p_0$, of finding an empty volume, $v$, or a
cavity of a given size and shape in water, \begin{equation}
\Delta\mu^{ex}=-kT \ln p_0.  \label{eqn:ktlnp0} \end{equation} We
calculate $p_0$ by considering the probabilities, $p_n$, of observing
exactly $n$ solvent centers in the cavity.  The $p_n$ are predicted by
maximizing an information entropy \cite{Hummer:96,berne:96}, subject to
the constraints of available information.  The experimentally accessible
first and second moments of the number of solvent centers in the cavity
region constitute the generally available information.  This procedure
yields here the distribution $p_n=\exp(\lambda_0+\lambda_1 n+\lambda_2
n^2)$ where $\lambda_0$, $\lambda_1$, and $\lambda_2$ are Lagrange
multipliers.  The required moments are obtained from the number density
$\rho$ and oxygen-oxygen radial distribution function $g(r)$ by
\begin{eqnarray} \langle n \rangle & = & \rho v, \\ \langle n(n-1)
\rangle & = & \rho^2\int_{v}d{\rm{\bf r}} \int_{v}d{\rm{\bf
r'}}g(|{\rm{\bf r} -{\bf r'}}|).  \end{eqnarray} That this is an
accurate model for the circumstances considered here has been explicitly
verified \cite{Hummer:96,berne:96}.

Figure \ref{fig:chempot} shows the calculated $\Delta\mu^{ex}$ for
spherical solutes as a function of temperature along the saturation
curve of liquid water.  For comparison, chemical potentials calculated
directly from simulation using test particle insertions \cite {Pratt:92}
are also shown.  The quantitative agreement between the two methods is
excellent over the entire temperature range.  The chemical potential
increases with temperature past 400 K but eventually decreases.  The
maximum in chemical potential occurs at about the same temperature in
each case.  These curves have the same shape as experimental ones
\cite{Harvey:91} for inert gases dissolved in water but shifted upward
due to the use of a hard sphere model.

Entropies calculated as the temperature derivative of $\Delta\mu^{ex}$
along the saturation curve are shown in Figure \ref{fig:entropies}.  As
expected, these entropies are large and negative at room temperature,
and increase with temperature.  The entropies of hydration for these
solutes converge in a temperature region about 400~K, close to the
temperature at which they are zero.  The observed entropy convergence
for transfer of simple nonpolar species from the dilute gas to water
\cite {Harvey:91} is similar.  Since water densities and O-O radial
distribution functions constitute the only input to our model, the
behavior shown in Figures \ref{fig:chempot} and \ref{fig:entropies} must
arise from these properties of liquid water.  We therefore examine the
connection between these properties and the solvation free energies.

A simplification of the two moment model gives an explicit expression
connecting the chemical potential to the density and density
fluctuations of liquid water:  \begin{equation} \Delta\mu^{ex} \approx
T\rho^2 \{ { k{v^2}/{ 2\sigma^2}}\} +T\{k\ln(2\pi\sigma^2)/2\}.
\label{eqn:loggauss}\end{equation} This is obtained from the Gaussian
estimate $p_n \approx $ $\exp (- {\delta n ^2}/{2\sigma^2} ) /{\sqrt {2
\pi \sigma^2}} $ where $\delta n = n - \langle n \rangle$, and $\sigma^2
= \langle \delta n^2 \rangle $.  The Gaussian formula is consistent with
the Pratt-Chandler \cite{Pratt:77} theory.  See
\cite{Hummer:96,berne:96,Chandler:93}.  Since the second term of eqn.
\ref{eqn:loggauss} is smaller than the first and is only logarithmically
sensitive to the size of the solute, this relationship says physically
that the solvation free energy may be lowered by decreasing the density
or the temperature of the solvent [the $T\rho^2$ factor]; or by
enhancing the ability of the solvent to open cavities of a size
necessary to accommodate the solute [the $\sigma^2$ factor in the first
term].  Along the saturation curve in Figure \ref{fig:chempot}, the
combination $T\rho^2(T)$ exhibits a nonmonotonic temperature dependence.
Surprisingly, $\sigma^2(T,v)$ has a negligible dependence on the
temperature over that range, so that \begin{equation} \Delta \mu^{ex}
\approx T\rho^2(T)x(v)+Ty(v).  \label{eqn:twoterm}\end{equation} The
quantities $x(v)$ and $ y(v)$, defined by the correspondence between
eqns \ref{eqn:loggauss} and \ref{eqn:twoterm}, depend only on the
excluded volume of the solute, not on the temperature.  Because the $T$
and $v$ dependence is important in both terms of eqn.  \ref{eqn:twoterm}
we require a generalization of the arguments \cite{Baldwin:86,Lee:91}
based upon the hypothesis of convergence at zero entropy.  Discussion of
the more general argument is illustrated by the schematic Figure
\ref{fig:schematic}.  To begin, note [Figure \ref{fig:schematic}(a)]
that if $Ty(v)$ were neglected in eqn.  \ref{eqn:twoterm} a precise
convergence at zero entropy would be observed; this follows from
previous arguments \cite{Baldwin:86,Lee:91}.  But $Ty(v)$ {\it is\/}
present, in fact, and adds $-k\ln(2\pi\sigma^2)/2$ to the entropies.
Since this contribution varies only logarithmically with $v$, it
provides [Figure \ref{fig:schematic}(b)] a downward shift, approximately
the same for each entropy curve.  An accurate convergence is obtained at
the entropy $-k\ln(2\pi\sigma^2)/2$.  Finally, if we include the
neglected temperature dependences by obtaining the temperature
derivatives of eqn.  \ref{eqn:ktlnp0}, the convergence is further
shifted to lower temperatures and entropies and is further blurred as
shown in Figure \ref{fig:schematic}(c).

In contrast, for nonpolar liquid solvents we anticipate a stronger
temperature dependence for $\sigma^2(T,v)$ in addition to the
nonmonotonic temperature dependence of $T\rho^2(T)$.  The temperature
dependences of both $T\rho^2(T)$ and $\sigma^2(T,v)$ would be needed to
describe the variation of the chemical potential along the coexistence
curve.  Eqn.  \ref{eqn:twoterm} would not apply and entropy convergence
would not be expected.  The macroscopic expression of this contrasting
behavior is the {\it relative\/} temperature insensitivity of the water
isothermal compressibility compared to hydrocarbon liquids; in
comparison to benzene, normal alkanes, and carbon tetrachloride, the
isothermal compressibility of liquid water varies only weakly along the
saturation curve up to 450 K \cite {Rowlinson:86}.

In conclusion, the temperature maximum in chemical potentials for
hydrophobic hydration and the entropy convergence both follow directly
from the weak temperature dependence of occupancy fluctuations
$\sigma^2(T,v)=\langle \delta n^2 \rangle$ for solute excluded volumes
in water.  Further, the value of the entropy at convergence is {\it
negative.\/} Interpretations of entropic contributions to protein
folding should account for this result.

The excellent agreement shown in Figure \ref{fig:chempot} between the
test particle insertion results and those obtained from the information
theory model demonstrates that a Gaussian distribution adequately
represents molecular-scale density fluctuations in liquid water.  This
ability to relate physical features of water structure to the observed
thermodynamics of hydrophobic hydration holds promise for extending the
information theory modeling to proteins and protein-ligand complexes in
aqueous solutions.

\begin{figure} \caption{ Excess chemical potentials $\Delta\mu^{ex}$ of
model hard sphere solutes of sizes roughly comparable to Ne, Ar, methane
(Me), and Xe as a function of temperature.  The hard sphere diameters
used were 2.8\AA, 3.1\AA, 3.3\AA, and 3.45\AA, respectively.  The hard
sphere diameter is then kept constant at various temperatures.  The
effects of more general solute-solvent interactions can be included
subsequently \protect\cite {Chandler:83}.  The lines indicate the
information theory model results and the symbols are the values
calculated using the test particle insertion method \protect\cite
{Pratt:92}.  Typical error bars in the test particle insertion method
are indicated for each solute.  The required second moments were
obtained from oxygen-oxygen radial distribution functions calculated
from Monte Carlo simulations of 256 simple point charge (SPC)
\protect\cite {spc-water} water molecules.  } \label{fig:chempot}
\end{figure}

\begin{figure} \caption{$ - \left( { \partial\Delta\mu^{ex}}
/{\partial T} \right)_{sat}$ along the saturation curve of liquid water
for model hard sphere solutes of sizes comparable to Ne, Ar, methane
(Me), and Xe as a function of temperature.  Note that the convergence
point would be more sharply defined if the Ne results were excluded.
Analysis of solubility data from different sources \protect\cite
{Harvey:91} locates the convergence temperature for Ne, Ar, methane, and
Xe in the neighborhood of 400 K.  This convergence region seems to move
systematically to lower temperatures and entropies as data on smaller
solutes are excluded, becoming more consistent with the values assumed
in protein unfolding experiments.  Closely examined, the width of the
entropy convergence region is several times 10~K for inert gas
solubilities. Additional,
 equation of state contributions to the standard solvation entropy
 are negligible: $|(\partial\mu^{\rm ex}/\partial T)_{\rm p} -
 (\partial\mu^{\rm ex}/\partial T)_{\rm sat}|<1$ and $<10$~(J/mol)/K
 for temperatures $T<450$ and $<550$~K, respectively.
}  \label{fig:entropies} \end{figure}

\begin{figure} \caption{$ - \left( { \partial\Delta\mu^{ex}}
/{\partial T} \right)_{sat}$ along the saturation curve of liquid
water, as in Figure \protect\ref{fig:entropies} but schematically.  (a)
Contribution to the entropy from the $T\rho^2(T)x(v)$ term of eqn.
\protect \ref{eqn:twoterm}.  This contribution dominates the eqn.
\protect \ref{eqn:twoterm}.  (b) Sum of the contributions from both the
terms in eqn.  \protect\ref{eqn:twoterm}.  The constancy of
$\sigma^2(T,v)$ with temperature is assumed.  For the hard sphere models
considered here the downward shift is about 7-9~J/mol/K.  (c) Entropies
calculated from eqn.  \protect\ref{eqn:ktlnp0} accounting for the
temperature dependence of $\sigma^2(T,v)$.  For the hard sphere model
solutes the leftward shift is roughly 40-50~K and the curves are shifted
downward by another 3-7~J/mol/K, approximately.  The total downward
shift of 10-16~J/mol/K is about a quarter of the entropy at room
temperature; see Figure \protect\ref{fig:entropies}.}
\label{fig:schematic} \end{figure} 
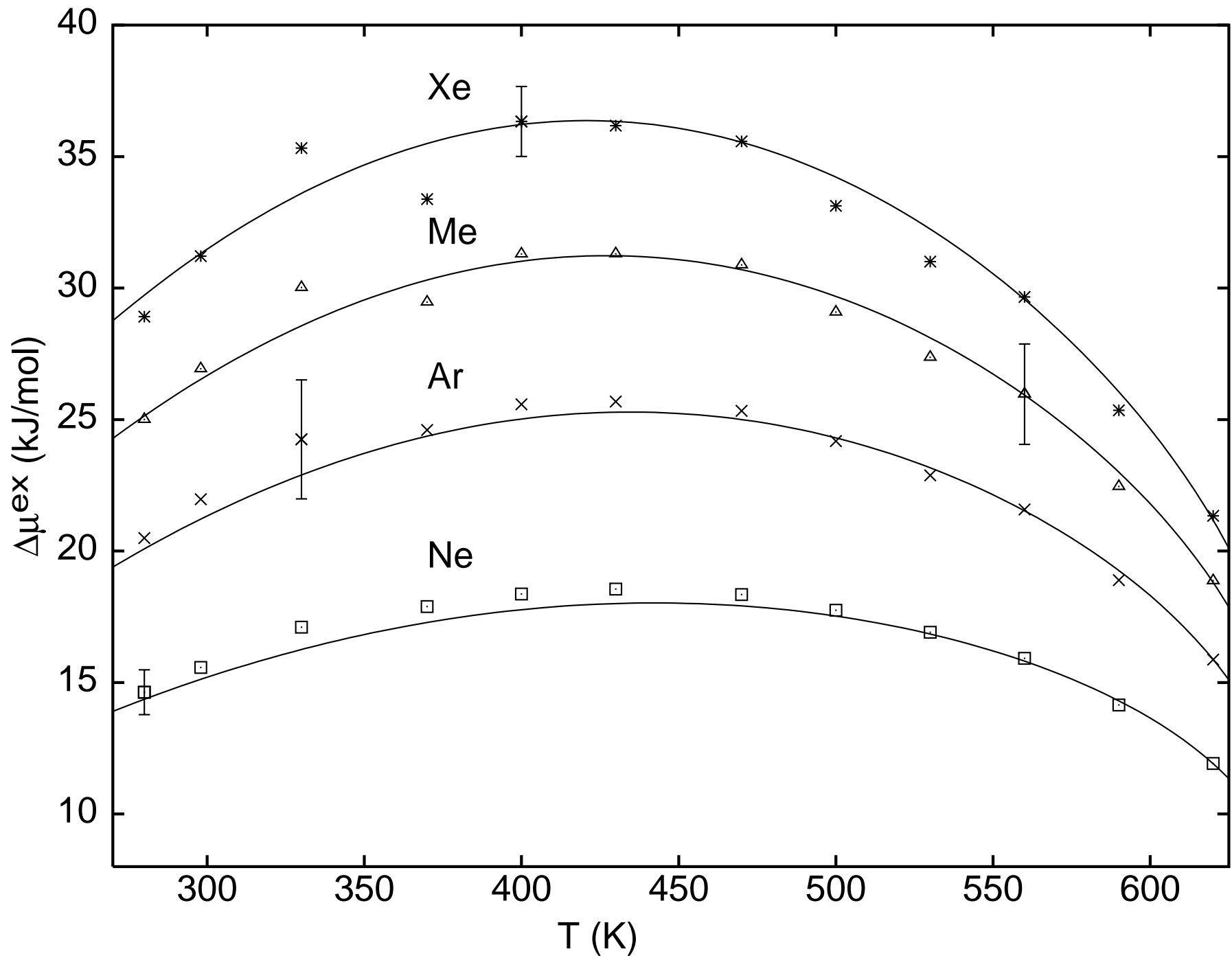
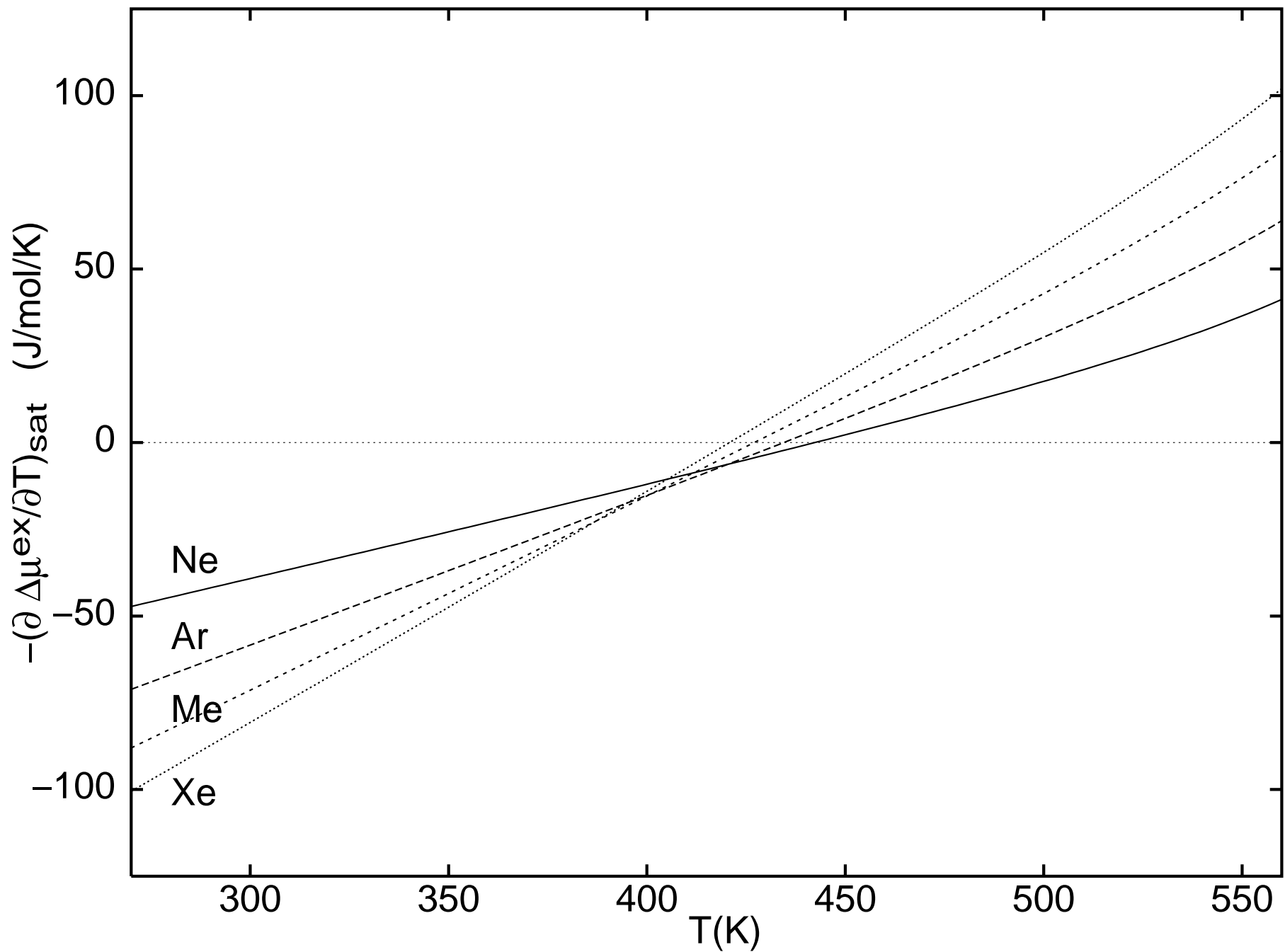
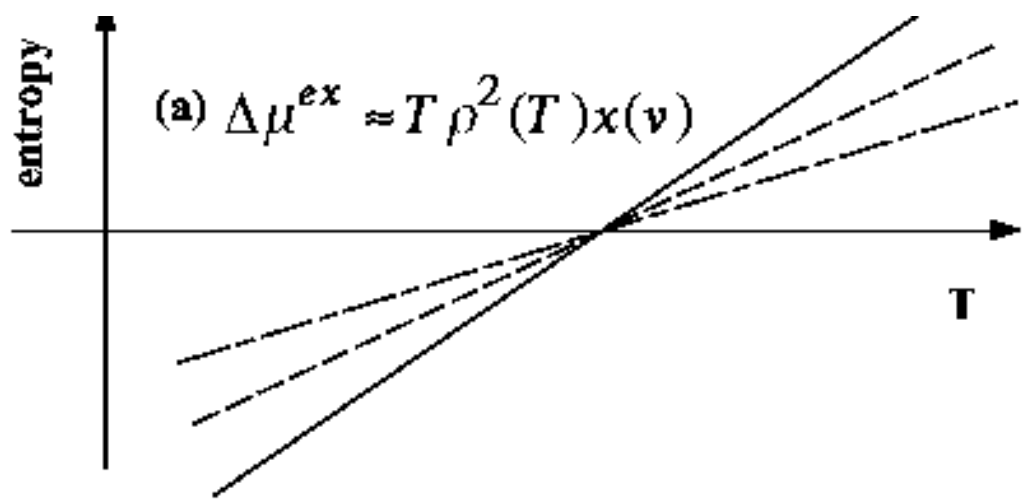
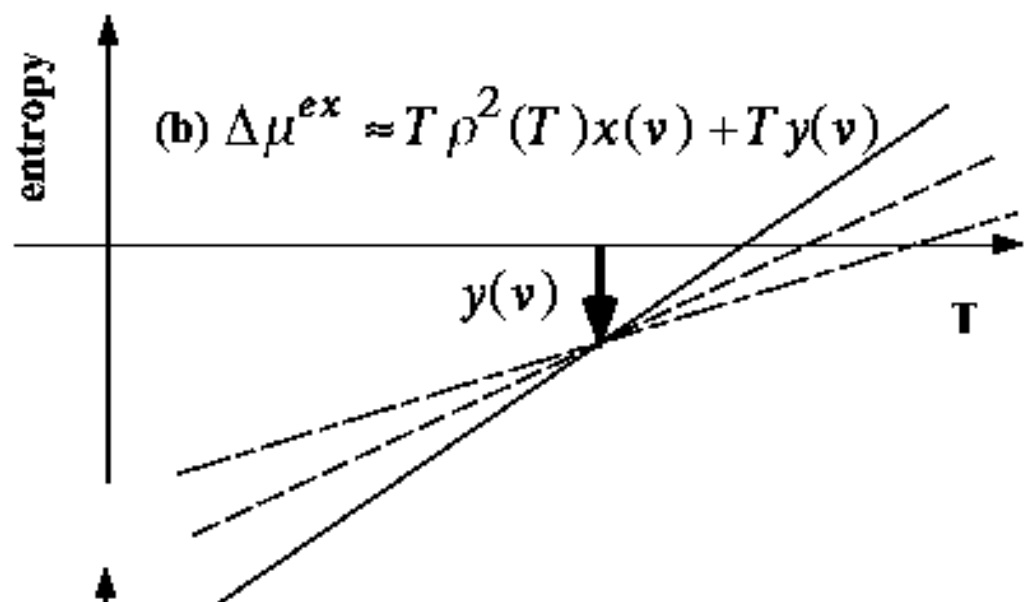
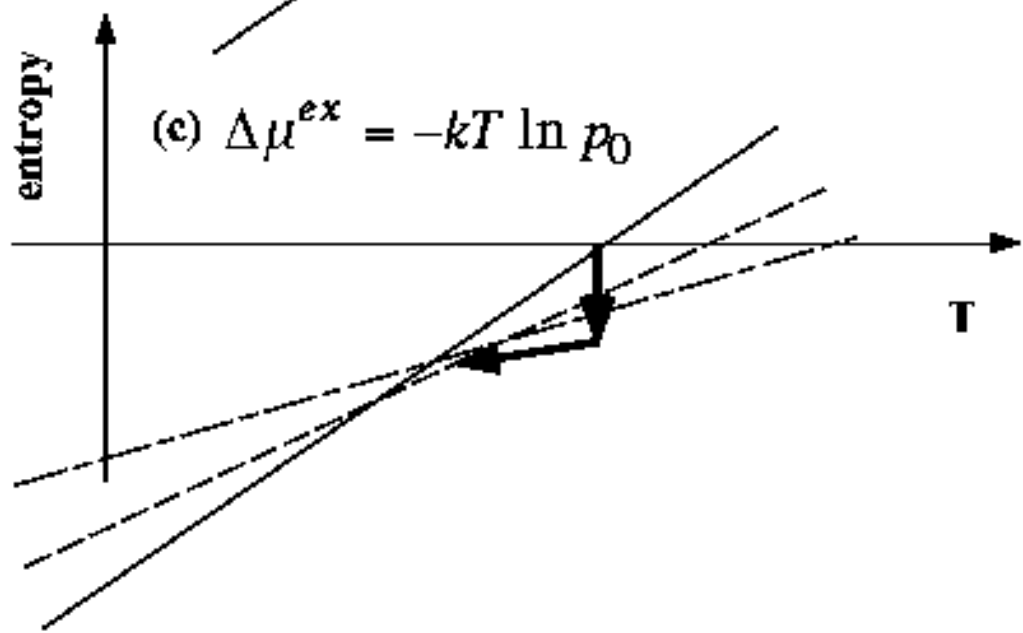
\end{document}